\documentstyle[preprint,aps]{revtex}
\begin {document}
\title{Bose-Einstein source of intermittency in hadronic interactions}

\author{Tadeusz Wibig}

\address{Experimental Physics Dept., University of \L \'{o}d\'{z}, \\
ul. Pomorska 149/153, PL-90-236 \L \'{o}d\'{z}, Poland}

\date{\today}

\draft

\maketitle

\begin{abstract}
The multi-particle Bose-Einstein correlations are the source of
''intermittency'' in high energy hadronic collisions. The power-law like
increase of factorial moments with decreasing bin size was obtained by
complete event weighing technique with gaussian approximation of space-time
particle emitting source shape. The value of source size parameter was
found to be higher than the common one fitted with the help of the
standard Handbury Brown-Twiss procedure.
\end{abstract}


The use of intensity interferometry to determine space-time sizes of the
particle emitting source is a well-established technique of high energy
physics. The standard methods based on Handbury Brown-Twiss
 \cite{HBT} (HBT) effect is to fit the Fourier transform
of the source space-time density to the two-particle correlation
function. The probability of finding one of the two emitted particles
with the momentum $p_1$ and the second with $p_2$ is given by

\begin{equation}
\label {eq1}
P_{\{ 12 \}} = \int {| \Psi ( x_1,x_2 ; p_1,p_2) |} ^2 \ \rho(x_1) \
\rho(x_2) \ d^4x_1 d^4x_2 ,
\end{equation}

\noindent where $x_1$ and $x_2$ are the four-positions of the emission
points each of them distributed in the ''source'' according to $\rho$.
If the particles are bosons a symmetrization in the amplitude $\Psi$
evaluation leads to the well-known formula:

\begin{equation}
\label{eq2}
P_{\{ 12 \} } \sim 1\  +\ {| {\cal F}_{12} |}^2 ,
\end{equation}

\noindent where the ${\cal F}_{12}$ can be related to the source
distribution by

\begin{equation}
\label{eq3}
{\cal F}_{12} = \int \ e^{( i q_{ij} x)}\ \rho(x)\ d^4x  ;\ \ \ \
  q_{ij}=p_i-p_j .
\end{equation}

There could be also other interpretations of ${\cal F}_{12}$. As for an
example the one given in Ref.\ \cite{andhof} derived on a basis of the
relativistic string fragmentation picture.

The particular choice of the source space-time distribution (or, more
general, the form of ${\cal F}_{12}$) leaves some degrees of freedom here,
but the results do not depend very much of that choice. The most popular
is the gaussian in space and exponential in time emission source shape.
However for the present work we choose the form of ${\cal F}_{12}$ which
is known as a gaussian parametrisation for its simplicity and because it
is Lorentz-invariant:

\begin{equation}
\label{eq4}
{\cal F}_{12} = e^{-(Q_{ij} R_0)^2/2 } ;\ \ \ \ Q^2_{ij} = -(p_i-p_j)^2
,
\end{equation}

\noindent which leads to the well-known formula for the two-particle
correlation function

\begin{equation}
\label{eq5}
C_2 (Q^2) = 1 \ + e^{-(Q R_0)^2 } .
\end{equation}

The $R_0$ can be still interpreted as a measure of space-time
extension of the emitting source.

The idea presented shortly above has extensively been used to analyze
different high energy physics data since first work
\cite{gold} by Goldhaber et al. Since that time many experimental and
theoretical efforts have been made. The different source shapes were
examined, some fine effects were predicted. Some difficulties were also
found in the interpretation of the source shape while the source is moving
very fast with respect to the laboratory system. However the main idea of
HBT effect remains unchanged.

In the mid-eighties due to the work of Bia\l as and Peschanski
\cite{biapa1} a new interest for the particle correlation has arisen. The
phenomenon called ''intermittency'' was found in the very small
phase-space bin size analysis. Since the first measurements the
experimentally available smallest bin size is reduced more than order of
magnitude but, what is even more important, new techniques to study fine
structures were developed. The ''intermittency'' of the particle creation
process seen by Bia\l as and Peschanski, which is in fact the fractal
(self-scaled) behaviour of the multiparticle correlation measures at the
very small phase-space scales, contradicts the standard Bose-Einstein
statistics driven description given by Eqs.\ (\ref {eq2}) and (\ref{eq3}).
The intermittent picture of hadronic creation was also inconsistent with
existing models of particle production (like e.g. LUND hadronization model).
The intermittent models like $\alpha$-model \cite {biapa1,biapa2},
geometrical branching model \cite {hwapan}, one-dimensional model of
intermittency by Dias de Deus \cite {DdD} were invented but none of them
achieved such a completeness and predictivity as high energy physics
standards (LUND or DPM-type models). On the other hand, the treatment
of the ''intermittency'' as a real new phenomenon was still not so
obvious. In Ref.\ \cite {Char1} different data sets were examined and as the
last conclusion it is stated that the intermittency is caused by
Bose-Einstein correlations {\it in addition} to a mechanism responsible
for the power-law behaviour, in Ref.\ \cite {carr1} authors claim
that the observed ''intermittent-like'' behaviour of moments of
multiplicity distributions can be understood as an effect driven by
quantum statistical properties of the particle emitting system and it
does not necessarily imply evidence for intermittency. The title of the
Ref. \cite {capp} ''Has intermittency been observed in multi-particle
production?'' is, to some extend, a good question still nowadays.

As it has been said, the existing data gives a possibility to study
intermittent (power-low) behaviour of factorial moments in more than two
decades wide phase-space distance measure (however it will be defined:
rapidity, momentum or four-momentum difference, box volume {\it etc.}). It
is clear now that the classical picture describing the HBT effect (Eq.
\ (\ref {eq2})) is not valid. The factorial moment analysis were performed
for ${\rm e^+e^-}$ annihilation into hadrons experiments, for hadronic
collisions ${\rm \pi^{+} /  K^{+}p}$ at ${\rm \sqrt{s} = 22 \
GeV}$ and at much higher energy ${\rm  \sqrt{s} = 630 \ GeV}$ ${\rm
p\overline p}$. There are also data from $p{\cal N}$ and ${\cal NN}$
experiments. All they shows more or less definite a power-law like
dependence on the bin size. However there is also very clear signal about
the like and unlike charge difference of the correlation strength which
suggests its Bose-Einstein origin. The possibility to achieve an
agreement between those two, on the first sight contradicting,
experimental facts will be discussed in the frame of common quantum
physics.

It should be remembered that Eq.\ (\ref {eq2}) was obtained in the case
when only two particles were emitted from the source. That situation is
of course different when one has got to do with multi-particle source
\cite {Zajc}. In some particular cases (when there are really a small
number of particles emitted in the large phase-space volume) the
two-particle correlator given by Eq.\ (\ref {eq2}) still can be used as at
least a first approximation. But when one wants to look closely at the
high multiplicity events or to study multiparticle correlations
Eq.\ (\ref {eq2}) has to be modified.

When n identical bosons are emitted the probability of the particular
momenta configuration ${\rm \{ p_i\} }$ is given by:

\begin{equation}
\label{eq6}
{P_{\{ n \}} } \sim {\sum_{\sigma} {\cal F}_{1 \sigma (1)}\
{\cal F}_{2 \sigma (2)}...\ {\cal F}_{n \sigma (n)} } ,
\end{equation}

\noindent where $\sigma$ is a permutation of a sequence
${\rm \{ 1, 2, ..., n \}}$, ${\rm \sigma (i)}$ is the i-th element
of this permutation and the sum is over all n! permutations.

To see what real difference is introduced by such complete treatment the
two-particle correlation function like that in the Eq.\ (\ref{eq2}) in
the case of three particle emitting source is written explicit below:

\begin{equation}
\label{eq7}
{P_{3}} \sim {1 \ + \ { | { \cal F}_{12} | }^2\ + \ {| {\cal F}_{13}
|}^2\ +
 \ {|{ \cal F}_{23}|}^2 \ + \ 2 \
| {\cal F}_{12} |\ | {\cal F}_{13}|\ |{\cal F}_{23}|} .
\end{equation}

If all three particles are very close to each other the statistical
weight of such events tends to ${\rm 6 = n!}$. The limit for two-particle
correlator in n-particle emitting source is n! not 2 like it comes form
Eq.\ (\ref{eq2}). The same limit was obtained in Ref. \cite {carr1} but
it was interpreted as a limit of n-th factorial moment. The multiplicity
distribution in the very small phase-space bin tends to the geometrical one
which, on the other side, can be treated as Bose-Einstein statistics
driven multiplicity distribution while ${\rm n \rightarrow \infty }$,
${\rm \delta \rightarrow 0}$ with ${\rm n \delta = const}$.

Quite different approach to Bose-Einstein phenomenon is discussed in
Ref.\ \cite{losjo}. The authors argued for the local nature of the
Bose-Einstein effect. In general, their treatments leads to the weighing
procedure with the event probability proportional to:

\begin{equation}
\label{eq8}
{P_{\{ n \}} '} \sim {\sum_{\rm all\ pairs} ( \ 1 \ + \ {|{{\cal
F}'}_{ij}|}^2)} .
\end{equation}

The definition of ${\cal F}'$ in Eq.\ (\ref{eq8}) is not given by
Eq.\ (\ref{eq4}) but is based on string fragmentation picture. However
the difference is rather in the physical interpretation than in the
general behaviour. It should be noted that Eq.\ (\ref{eq8})
overestimates the very close particle limit. It is there equal to
${\rm 2^{n(n-1)/2}}$. The arguments for such a treatment are discussed
in Ref.\ \cite {losjo} (similar attempt is presented in Ref.\ \cite{ab})
and will not be discussed here.
One of the arguments not given there but of the practical
importance is that the above idea can be easily incorporated into the
Monte-Carlo event generator. It was in fact done in the LUBOEI
subroutine which is a part of LUND hadronization scheme JETSET 7.3. The
general difference between the Ref.\ \cite {losjo} strategy and proposed
in the present paper is in the fact that sum in the Eq.\ (\ref{eq8}) is
performed over permutations of the particle ensemble in which only two
particles are exchanged (locality of Bose-Einstein interaction) while in
our treatment all event permutations can give a contribution to the event
weight (global Bose-Einstein approach). The importance of many particle
exchange contributions will be discussed later on.

The problem with complete weighing procedure is also a practical one. The
sum over n! elements can be performed easily for about of ten particles
or less. For higher multiplicities the calculation time rise
tremendously. But it is quite clear than for the two very distant
particle exchange the contribution coming from all permutations
concerning that particular exchange is negligible. The algorithm was
invented to omit all the negligible permutations and calculations of the
weights according to Eq.\ (\ref{eq6}) became possible also for larger
multiplicities. In the present paper only the data from NA22
experiment will be analyzed. The mean charged particle multiplicity is
of order of 8 and the largest like type boson multiplicity (in one chain,
as will be discussed later) in the sample of about 500000 our
Monte-Carlo generated events does not exceed 15.

To study the influence of Bose-Einstein weighing method on the shape of
the fluctuations in small bins the sample of events in the ''world of
absence of Bose-Einstein correlation'' is needed. There is a number of
Monte-Carlo generators which can be used to get this. In the present work
the one called Geometrical Two-Chain was used. It is described in
details in Refs.\ \cite {ws1}. The advantage of that generator is the
minimum of correlations introduced there. The ones existing are due to
the conservation requirements (charge, barion number, strangeness,
momentum and energy), the resonance production and the large scale
clustering due to chain mass distribution in the model. There are
also correlations connected with the hadronization procedure adapted:
the transverse momentum is conserved locally in the fragmenting chains
so the subsequent hadrons incline to have the negatively correlated momenta
perpendicular to the interaction axis. Our chain fragmentation picture
leads also to ordering in rapidity of subsequently produced hadrons. All
that features are present in most of the models working on the partonic
level. The last but very well seen specially for large bin sizes
is a contribution related to non-poissonian multiplicity distribution
in the multiparticle production.

The main interaction characteristics are very well reproducible by the
generator as it was shown in Refs.\ \cite {ws1}.

About 500000 of non-single-diffractive events for $ \pi ^{+}$ and
${\rm K^{+}}$ interactions with proton at laboratory momentum of
${\rm 250 \ GeV/c}$ were generated and combined to get the reference
sample without Bose-Einstein correlations included.
Then for each event the weight was calculated according to
Eqs.\ (\ref{eq4}) and (\ref{eq6}). In principle the Bose-Einstein weighing
procedure could change the multiplicity distribution (what was one of the
argument against global treatment of Bose-Einstein correlation in Ref.\
\cite {losjo} ). To avoid this the weights were renormalized to get the
average value of the weights for n identical bosons equal to 1 and these
were used afterward. The detail comparison with the experimental data leads
to the conclusion that if the Bose-Einstein symmetrization were performed
for the whole events then the correlations are too strong for very small bin
sizes. In our model there is only one parameter to be adjusted, correlation
radius $R_0$, while in the standard HBT procedure there is also the incoherence
parameter which allows to make softer the correlation strength. In the
Geometrical Two-Chain model particles are produced by the fragmentation of
two well-defined chains so there is a natural subdivision of all secondary
particles to two distinct classes. To make the correlation weaker there is
a possibility to symmetrize amplitudes $\Psi$ not over all particle
exchanges but only over the exchanges of the particles produced from the
same chain.

The very convenient variable to study the two-particle correlation is the
differential form of the second factorial moment as it was used in
Ref.\ \cite {agab1}. The definition using density integral method \cite{lipa}
is:

\begin{equation}
\label{eq9}
D_2 (Q^2) = {1 \over { Norm}} \ 2 \ \sum_{i<j}
 \ {\Theta (Q^2-Q^2_{ij})} \times {\Theta (Q^2_{ij}-Q^2+\delta )}   ,
\end{equation}

\noindent were $\Theta$ is the Heaviside unit step-function and
$Norm$ is a normalization term defined by the
so-called ''mixed events'' technique. The particles used for the
normalization were chosen randomly from the all event ''pool'' of the
large number of generated interactions, ensuring that they belongs to
different real events. To avoid in the reference sample the correlations
due to non-poissonian multiplicity distribution in hadronic interactions
the multiplicity in the mixed events were taken from Poisson distribution
with the average value the same as in M-C generated events.

In the particular NA22 experimental data which we want to compare with
the rapidity cut ${\rm |y|<2}$ has been used. Thus in all the calculations
the same cut is applied. In the experimental procedure there was also
not possible, in general, to determined the particle masses so all the
particles (except low energy proton and very energetic particles in ${\rm
K^{+}}$ induced interactions ${\rm p_{lab} > 150 \ GeV/c}$) were treated
as pion. The same procedure has been used in our analysis of the
Monte-Carlo events. The experimental accuracy of particle four momentum
difference determination described in Ref.\ \cite {kitt1} was taken into
account in the calculations as well. The calculations od $D_2$ was
performed for all charged particles as well as for like and unlike
charge combinations.
The results are presented in Fig.\ \ref{fig1} by the solid line. The
remained correlations produced in the Geometrical Two-Chains model, which
were indicated above, leads to the outcome depicted by the dotted line.
It represents the result of the correlation calculations without
Bose-Einstein weighing.

It is seen that the power-low-like behaviour of ${D_2(q^2)}$ is
quite well reproduced by our weighing method.
The small overestimation of the unlike particle correlator at four-momentum
difference of about $Q^2 \sim 10^{-2} \div 10^{-1} {\rm (GeV/c)^2}$
is a consequence of the strict ordering in rapidity of the chain
fragmentation products which introduces always between close (in rapidity)
like type charged hadron the one with the opposite sign. The four-momentum
difference of that unlike charged pairs is determined by transverse momentum
distributions so the effect does not influence the very small bin size
analysis.

However, the main argument for intermittency comes from the analysis of
the higher multiplicity correlation measurements. To study this effects
the correlation measures have to be defined for three- and more particle
systems. The most commonly used variables are the factorial moments. For
practical purposes the best method of factorial moment calculations is
again the one proposed in Ref.\ \cite{lipa} density integral method.

\begin{equation}
\label{eq10}
F_q(Q^2)= {1 \over { Norm}} \ q!\  \sum_{i(1) <i(2) <...< i(q) }
   \ \        \prod_{{\rm all\ pairs }\ (i(k1), i(k2))}
\Theta (Q^2-Q^2_{i(k1) i(k2)}) ,
\end{equation}

\noindent with the normalization by mixed event technique again.
Results of our calculations are presented in Fig.\ \ref{fig2}. The
power-law like increase of factorial moments with decreasing bin size
is again quite well reproduced in the whole range of $Q^2$ measured
experimentally.

In Figs.\ \ref{fig1} and \ref{fig2} the results of event weighing defined
by Eqs.\ (\ref{eq4}) and (\ref{eq6}) with the sum over permutations with
only one particle pair exchange are also presented (by the dashed line).
As it is seen the effect (for the same value of $R_0$ parameter in
Eq.\ (\ref{eq4}) ) is much weaker. This illustrate the importance
of global treatment of Bose-Einstein correlation. The introduction to
the sum of the weights very many relatively small terms leads to really
great increase of the effect.

To reproduce the shape of ${D_2(Q^2)}$ and ${ F_2(Q^2)}$ dependencies
measured by NA22 experiment the value of the parameter $R_0$ in
Eq.\ (\ref{eq4}) had to be adjusted. The large statistical fluctuations
of the weights influence the estimation of source size parameter so the
accuracy achieved is not higher that ${\rm 10 \% }$. In Ref.\ \cite{agab1}
the source size was found using the standard technique of HBT effect
(Eq.\ (\ref{eq5}) ). The value found there was ${\rm ( 0.82 \pm 0.02 )\ fm}$.
Our complete weighing procedure gives stronger correlations ( even after
weights renormalization ) so the value of $R_0$ used to obtain the results
given in Figs.\ \ref{fig1} and \ref{fig2} is about ${\rm  50 \% }$ higher
what gives the source radius of about ${\rm 1.25 \ fm}$
in the gaussian approximation (Eq.\ (\ref{eq4}) ) interpretation.

To summarize, the importance of the global treatment of Bose-Einstein
correlation has been shown. The symmetrization over all permutations
leads to the power-low like behaviour of factorial moments in the
four-momentum difference regions where they are measured experimentally.
The more detail analysis is in progress and the results will be presented
elsewhere.

\begin{figure}
\caption{ The differential second factorial moments for a) all charged,
b) like-charged and c) unlike-charged pairs as a function of
four-momentum difference. The data points are from NA22 experiment.
Solid line represents the result of our complete Bose-Einstein
weight method, dotted shows the correlations in used Geometrical
Two-Chain model. Dashed line is for the sum in Eq.\ (6) over only
one pair of boson exchanges.
\label{fig1}}
\end{figure}

\begin{figure}
\caption{ The factorial moments a) for negatives and b) for all
charged particles as a function of four-momentum difference.
The data points are from NA22 experiment.
\label{fig2}}
\end{figure}

\end{document}